\documentclass{ws-procs975x65}
\def\ol{\overline}
\def\]{\right]}
\def\be{\begin{eqnarray}}
\def\ee{\end{eqnarray}}
\def\nn{\nonumber}
\def\({\left(}
\def\){\right)}
\def\bk#1{\langle#1\rangle}
\def\eq#1{(\ref{#1})}
\def\rd{\sqrt{2}}
\def\rt{\sqrt{3}}
\def\rs{\sqrt{6}}
\def\bp{\begin{pmatrix}}
\def\ep{\end{pmatrix}}
\def\diag{{\rm diag}}
\def\G{{\cal G}}
\def\o{\omega}
\def\bk#1{\langle#1\rangle}
\def\ul{\underline}

\begin{document}

\title{Horizontal Symmetry: Bottom Up and Top Down}

\author{C.S. Lam}

\address{McGill University and the U. of British Columbia, Canada\\}

\begin{abstract}
A group-theoretical connection between horizontal symmetry $\G$ and fermion mixing
is established, and applied to neutrino mixing. The group-theoretical approach is
consistent with a dynamical theory based on $U(1)\times \G$, but the dynamical
theory can be used to pick out the most stable mixing that purely group-theoretical considerations cannot.
A symmetry common to leptons
and quarks is also discussed. This higher symmetry picks $A_4$ over $S_4$ to be the preferred symmetry for leptons.
\end{abstract}


\bodymatter

\section{Introduction}
When neutron was discovered by Chadwick in 1932, Heisenberg came up with a new quantum number, the isotopic spin, to distinguish it
from the proton. The associated  Noether symmetry $SU(2)$ valid for strong interactions
tells us that their nuclear forces are essentially the same. Now that
we have three generations of quarks and three generation of leptons, all with the same Standard Model interactions, it seems natural 
that there would be a new quantum number and a new symmetry for them. This hypothetical symmetry is known as a horizontal
symmetry, a family symmetry, or a generation symmetry. 

Unlike protons and neutrons, which have nearly the same mass, the masses of the three generations of fermions are
vastly different. Moreover, they mix. This suggests that if a horizontal symmetry exists, it has to be severely broken, presumably
spontaneously, a fact which makes the identification of symmetry very difficult. There is however hope if the breaking leaves
behind some unbroken residual symmetry that can be traced. In that case,  group
property of the symmetry can be invoked to reconstruct the unbroken horizontal symmetry from the residual symmetry.

We will show in the next section that residual symmetries are present and located in the mixing matrix. In the case when neutrino mixing is
given by the tribimaximal mixing matrix \cite{HPS}, the minimal horizontal symmetry group so obtained is $S_4$, the permutation group
of four objects. 

\section{Group Theory of Mixing}
In this section we show how mixing determines the residual symmetry, and from there the horizontal symmetry.  We also show how
mixing can be obtained from a given horizontal group, 
using only symmetry arguments with no dynamical input. We will concentrate on leptonic
mixing which produces a finite non-abelian symmetry group. The same procedure can in principle be used to study quark mixing,
but no symmetry smaller than $SU(3)$ is obtained in that way. For more details of this approach, see Refs.~[2] to [4].

Let $M_e$ be the charged-lepton mass matrix, $M_\nu$ the neutrino Dirac mass matrix, and $M_R$ the heavy Majorana
mass matrix in a type-I seesaw model. Since a mixing matrix  refers to the mixing of left-handed fermions, we can eliminate
any reference to the right-handed fermions by considering the effective mass matrices $\ol M_e=M_e^\dagger M_e=\ol M_e^\dagger$ 
for charged leptons,
and $\ol M_\nu^T=M_\nu M_R^{-1}M_\nu=\ol M_\nu^T$ for neutrinos, which connect only left-handed fermions
with one another.
 The fact that $\ol M_\nu$ is a symmetric rather than hermitian like $\ol M_e$
is a reflection that neutrinos are (assumed to be) Majorana particles.

The PMNS neutrino mixing matrix $U$ is the matrix that renders $U^T\ol M_\nu U$ diagonal in the basis where $\ol M_e$ is diagonal.

Suppose $F$ is a symmetry for left-handed charged-leptons and $G$ a symmetry for left-handed neutrinos. 
That means, $F,G$ are unitary matrices obeying $F^\dagger \ol M_eF=\ol M_e$ and $G^T\ol M_\nu G=\ol M_\nu$.
We shall assume both of them to have a unit determinant so that they belong to $SU(3)$. 
It can be shown from the Majorana character $\ol M_\nu=\ol M_\nu^T$ that $G^2=1$. That means $G$ has two $-1$ eigenvalues and one
$+1$ eigenvalues. As a result, whatever the PMNS mixing matrix $U$ is, there are exactly three residual symmetry operators $G$ given by
\be
{\scriptsize G_1=U\diag(1,-1,-1)U^\dagger,\ G_2=U\diag(-1,1,-1)U^\dagger,\ G_3=U\diag(-1,-1,1)U^\dagger}.\nn\ee
It is easy to see that the product of any two of them equals to the third, so they are the non-unit elements of an abstract $Z_2\times Z_2$ group. 

The residual symmetry $F$ in the charged-lepton sector is different. Since $\ol M_e$ is diagonal and non-degenerate, all that is required
 is for $F$ to be diagonal and unitary. It can satisfy $F^n=1$ for any $n$. We would however limit it to non-degenerate
matrices, with three distinct eigenvalues, so that in the basis where $F$ is diagonal, $\ol M_e$ is forced to be diagonal as well.
In principle there are still an infinite number of these symmetry operators, forming an abstract group which is the direct product of any number
of cyclic groups.

\subsection{From $U$ to $\G$}

Any group $\G$ generated by $G_1, G_2, G_3$ and at least one $F$ is a possible horizontal symmetry group of the left-handed
fermions from which mixing $U$ 
can be obtained. The broken symmetries are those elements in $\G$ not equal to $F$ and $G_i$.
If we want $\G$ to be minimal,  we pick only one $F$, and the smallest possible $F$ at that. That would be
$F=\diag(1,\o,\o^2):=F_3$ where $\o=e^{2\pi i/3}$. 

We will use the symbol $\{X_1,X_2,\cdots, X_m\}$ to denote the group generated by the matrices $X_1$ to $X_n$. In that notation,
the horizontal symmetry group is $\G=\{F,G_1,G_2,G_3\}$.

There are a few general things we can say about $\G$.
To have a non-trivial mixing, $G_i$ must not be diagonal when $F$ is, hence $\G$ is a non-abelian group. Since it contains
a subgroup generated by the $G_i$'s, its order must be a multiple of 4. Moreover, for almost all $U$'s, the order of the group $\G$
would turn out to be infinite; the symmetry operators $G_1, G_2, G_3$ have to be just right to produce a finite group $\G$. One of
these very special $U$'s turns out to be the tribimaximal mixing matrix discussed below.

Let us now apply this general formalism to neutrino mixing observed experimentally. To within one standard deviation, the mixing matrix 
is consistent with the `tribimaximal matrix' (TBM) \cite{HPS}
\be U_{TBM}={1\over \rs}\bp 2&\rd&0\cr -1&\rd&\rt\cr -1&\rd&-\rt\ep,\label{TBM}\ee
whose mixing angles $\theta_{ij}$ and CP phase $\delta$ are given by
\be P:=\langle\sin\theta_{12},\sin\theta_{23},\sin\theta_{13}e^{-i\delta}\rangle=\langle {1\over \rt},{1\over\rd},0\rangle.\label{PTBM}\ee
The residual symmetries $G_i$ derived from this $U$ are
\be
G_{1}={1\over 3}\bp 1&-2&-2\cr -2&-2&1\cr -2&1&-2\ep,\ G_{2}={1\over 3}\bp -1&2&2\cr 2&-1&2\cr 2&2&-1\ep,\ 
G_{3}=-\bp 1&0&0\cr
	0&0&1\cr 0&1&0\cr\ep.\label{GGG}\ee
Defining an {\it invariant eigenvector} to be an eigenvector with eigenvalue $+1$, \eq{GGG} simply says that
the normalized invariant eigenvector of $G_i$
is the $i$th column of the mixing matrix $U$. Thus in the case of TBM, $G_3$ is responsible for bimaximal mixing in the 
third column of $U_{TBM}$,
and $G_2$ is responsible for trimaximal mixing in the second column.

Choosing $F=\diag(1,\o,\o^2):=F_3$ to be the residual symmetry in the charged-lepton sector, the horizontal symmetry
group generated by $F$ and the these three $G_i$'s turns out to be the symmetric group $S_4$. 

If $F$ is not given by $F_3$, then the group $\G$ is not $S_4$. However, it can be shown that as long as $\G$
is a finite subgroup of $SU(3)$,  it always contains $S_4$ as a subgroup \cite{LAM1}.

Let us look at another example to further illustrate the general procedure. This example is unphysical, for its mixing angles
and CP phase  (see \eq{PTBM} for definition) are $P=\langle{1\over\rd},{1\over\rd},{e^{-\pi i/2}\over\rt}\rangle$,
far from the experimental values. Nevertheless, this example is useful in illustrating a couple of other things.

The mixing matrix in this example is
\be
U_{CW}={1\over\rt}\bp 1&1&1\cr \o&1&\o^2\cr \o^2&1&\o\cr\ep,\label{ucw}\ee
and the resulting residual symmetries $G_i$ computed from $U_{CW}$ are
\be
G'_{1}={1\over 3}\bp -1&2\o&2\o^2\cr 2\o^2&-1&2\o\cr 2\o&2\o^2&-1\ep,\ G_{2}={1\over 3}\bp -1&2&2\cr 2&-1&2\cr 2&2&-1\ep,\ 
G'_{3}={1\over 3}\bp -1&2\o^2&2\o\cr
	2\o&-1&2\o^2\cr 2\o^2&2\o&-1\cr\ep.\label{GGGA4}\ee
Note that $G_2$ is identical in \eq{GGG} and \eq{GGGA4}, but $G_1$ and $G_3$ are different, so a prime
is put in \eq{GGGA4} to tell them apart. The group $\G=\{F_3, G_1', G_2,G'_3\}$ is now $A_4$, the subgroup
of $S_4$ consisting of even permutations of four objects.

It turns out that $\{F_3,G_1\}$ is already $S_4$ and $\{F_3, G_2\}$ is already $A_4$. The physical significance of
this observation is explained below.

Remember that the normalized invariant eigenvector of $G_i$ is the $i$th column of the mixing matrix $U$. If $G_2$ is given, then the
second column of $U$ is fixed. The symmetry group $\G$ of $U$ depends on what we choose for the first and third columns, equivalently,
what $G_1$ and $G_3$ are. Most of the time the resulting
group $\G$ is very large, of infinite order. If we look at the other end of the spectrum, and ask how can
they be chosen so that the resulting group $\G$ is the smallest, namely $A_4=\{F_3,G_2\}$, then the answer is that they must be
$G_1'$ and $G_3'$ of \eq{GGGA4}. If we ask how can they be chosen to obtain the next smallest group, namely $S_4$, then the answer
is $G_1$ and $G_3$ of \eq{GGG}. Similarly, if the first column of the TBM matrix is given, then the smallest group $\G=S_4$
is obtained by choosing the second and third columns of $U$ to be those of TBM. We can get a larger group, such as $S_5$, by choosing
these two columns differently, but most of the time we will end up with a group of infinite order for a random choice of these two columns.

\subsection{From $\G$ to $U$}
So far we have discussed how to obtain the unbroken horizontal symmetry group $\G$ from a mixing matrix $U$. The connection also
works in the opposite direction, getting $U$ from a given $\G$. This can be done in the following way \cite{LAM2}. 

First, identify all possible pairs of mutually commuting order-2 elements in $\G$. These would be candidates for $G_1$ and $G_2$, with
$G_3$ given by $G_3=G_1G_2$. Next, pick a $F$ from any  other element with an order $\ge 3$, provided
its eigenvalues are non-degenerate. Go to a 3-dimensional irreducible representation in which ${\rm det}(G_i)=+1$,  in the basis
 where $F$ is diagonal. Pick out the invariant eigenvectors $u_1, u_2, u_3$
of $G_1, G_2, G_3$ respectively. Then the three columns of the mixing matrix $U$ are simply $u_1, u_2$, and $u_3$.

It is clear that there is no way for group theory to tell which column is which, and similarly which row is which. Hence the $U$
determined this way is unique 
only up to possible re-shuffling of rows and columns. Moreover, since the Majorana phases are not presently known,
we can identify two $U$'s differed only by row and column phases. We will consider two $U$'s {\it equivalent} if they
differ only by row and column phases, and row and column re-shuffling.

One might think that there are so many ways to choose the pair $(G_1, G_2)$ and the  element $F$, that the number of $U$'s 
emerging would be so large to become unmanageable. 
It turns out that this is not the case as most of them are equivalent, leaving behind only very few inequivalent ones, at least for 
small groups. For $\G=A_4$, the only inequivalent mixing  is given by \eq{ucw}. For $S_4$, there are two inequivalent
mixings, given by
the TBM in \eq{TBM}, and another one with $P=\langle{1\over\rd},{1\over\rd},0\rangle.$ For $A_5$, there are again only 2 inequivalent ones.
For more detail, please see Ref.~[3].

Sometimes we refer to this kind of mixing as {\it full mixing}, to distinguish it from a {\it partial mixing}, 
formed by one order-2 element $G$ and one order $\ge 3$ element $F$
 picked from $\G$, with a second order-2 operator $G'$ commuting with this $G$
picked arbitrarily, not necessarily from the group. If $\G=\{F, G\}$, then $\G'=\{F, G, G'\}\supseteq G$. If no adjective is attached,
then mixing means full mixing. For example, TBM can be obtained as a full mixing of $S_4$, or a partial mixing of $A_4$.

\section{Dynamical Theory of Mixing}
So far everything is derived from group-theoretical considerations, without any dynamical input.
What would we gain by incorporating dynamics into consideration? And, what kind of dynamics should we impose?

Let me discuss the second question first. The dynamics should of course be invariant under the symmetry group $\G$. 
We would also like the mixings derived from it to retain some memory of the group, namely, 
to have at least one residual symmetry each to be a member of $\G$, both in the charged-lepton sector and 
the neutrino sector. In other words, we would like the mixings deduced from the dynamics to be 
full or partial mixings of $\G$.
There is no a priori
guarantee that such a dynamics exists, but for many groups including $A_4$ and $S_4$, the dynamics can indeed be obtained
 by imposing an additional $U(1)$ symmetry,
as we shall discuss in the second subsection below. 

To implement this objective we must first figure out how to impose
a residual symmetry in a dynamical theory. That will be discussed in the first subsection below. 
Now the first question. With dynamics different mixings have different energies, 
so dynamics provides a mechanism to pick out the preferred inequivalent or partial mixing that has the lowest energy among them.

If dynamics can pick out a preferred mixing (or residual symmetries) from a group $\G$, then in a similar way it might be able to pick out a
preferred group $\G$ from a larger group. This possibility is discussed in the third subsection below. 

The lack of time prevents me from discussing any of these in great detail, so I will only summarize the results here,
and refer the interesting
readers to the published literature \cite{LAM2, LAM3}.

\subsection{Vacuum alignment}
A dynamical theory starts from a Lagrangian invariant under s horizontal symmetry group $\G$. Mass matrices from which mixing is derived
come from the Yukawa interactions after symmetry is broken spontaneously. 
The Higgs fields in the Yukawa terms carry Standard Model (SM) quantum
numbers and $\G$ quantum numbers. For simplicity we shall assume them to be compound fields, made up of products of the usual SM
Higgs fields  and {\it valon} fields carrying horizontal ($\G$) quantum numbers. The SM Higgs fields are horizontal singlets, and the
horizontal valon fields are SM singlets. This allows us to consider horizontal symmetry separately from the 
known SM symmetries.

The original Lagrangian is invariant under $\G$, hence if we carry out a simultaneous $\G$-transformation of the left-handed
fermions, the right-handed fermions, and the valons, then every Yukawa term remains unchanged. This is no longer true when the
valons acquire expectation values, thereby breaking $\G$. However, if the valon expectation value is an invariant eigenvector of 
some $g\in\G$, then the Yukawa terms containing this valon would remain invariant under $g$. Therefore, if $\phi$ is a valon
field in the charged-lepton sector such that $F\bk{\phi}=\bk{\phi}$, then $F$ is a residual symmetry of the charged lepton
mass matrix $M_e$. Similarly, if
$\chi$ is a valon field in the neutrino sector such that $G_i\bk{\chi}=\bk{\chi}$, then $G_i$ is a residual symmetry of
the neutrino mass matrices $M_\nu$ and $M_R$. If this $\bk{\chi}$ is the same for all three $i$, then the mixing computed from it
is a full mixing. If it is different for different $i$, then the mixing computed from it is a partial mixing. In that case full mixing is
obtained by taking $\bk{\chi}=0$, the only common invariant vector of all the $G_i$'s. 

In order to have enough tunable parameters to fit the masses, the original Lagrangian usually contains valons belonging to all the 
irreducible representations of $\G$. The condition above that expectation values are invariant eigenvectors must be true in all
irreducible representations. In case such an invariant eigenvector does not exist in some representation, then the expectation
value should be taken to be zero because a zero vector is always an invariant vector.

For illustrations and more details, see Ref.~[2].

\subsection{Valon dynamics of $U(1)\times \G$}
Expectation values are taken from the stationary points of a valon potential.
According to the last subsection, in order for the potential to produce only full and partial mixings of $\G$,
these points have to be invariant eigenvectors in every irreducible representation.

We will use $\G=A_4$ to sketch out what is involved. For detailed discussions and for other groups $\G$, see Ref.~[3].

 There are four irreducible representations in $A_4$,
 $\ul 1,\ \ul 1',\ \ul 1''$, and $\ul 3$. 
The expectation values in each of the one-dimensional
irreducible representations is either zero or not zero. In either case it is easy to write down a potential that produces the result. 
Thus, assuming
the potentials of different irreducible representations do not interact, the difficult ones 
are those involving multi-dimensional representations such as  $\ul 3$.

For \ul 3, the solution must be  either (0,0,0), or (1,0,0) which is
the invariant eigenvector of $F=F_3$, or $(1,\o,\o^2)$ which is the invariant eigenvector of $G_1'$, or (1,1,1) which is the invariant
eigenvector of $G_2$, or $(1,\o^2,\o)$ which is the invariant eigenvector of $G_3'$, or one of their equivalents.

Is there a generic potential of \ul 3 that has exactly these solutions? The answer is `yes', because 
a generic $U(1)\times A_4$ potential does indeed have these properties \cite{LAM2}.

This potential also tells us that the TBM mixing matrix \eq{TBM} obtained as a partial mixing of $A_4$ 
is energetically more favorable than the full mixing \eq{ucw}.

\subsection{$U(1)\times SO(3)$ dynamics}
Both $\G=S_4$ and $A_4$ are capable of producing TBM, so is the horizontal symmetry of leptons $S_4$, $A_4$, or something else? Since
both of them are subgroups of $SO(3)$, one might hope to use the dynamics of $SO(3)$ to see which 
of these finite subgroups is $SO(3)$ going to break down to, thereby picking out the preferred horizontal symmetry for leptons.
Unfortunately, it is known \cite{SO3} that no matter what the dynamics is, as long as the valons have an $SO(3)$ spin less than 3, neither
$S_4$ nor $A_4$ symmetry can be retained after the breakdown. We want valons to have spins less or equal to 2 because they
couple to two fermions, and being a triplet each fermion can have a horizontal spin of at most 1.

However, a generic $U(1)\times SO(3)$ valon potential can fulfill this mission. Depending on the
relative strengths of coupling constants, it breaks into one of three possible phases, one of which carries
an $A_4$ symmetry suitable for lepton
mixing. No phase containing any other non-abelian
symmetry such as $S_4$ exists. 
 The other two phases produce block-diagonal and hierarchical mass matrices that are capable of describing mass hierarchy and
Cabibbo mixing of quarks \cite{LAM3}. Thus this higher symmetry $U(1)\times SO(3)$ not only picks out $A_4$ over $S_4$, it also serves as an
approximate common
symmetry for quarks and leptons at some high energy scale.


\begin{thebibliography}{9}
\bibitem{HPS} P.F. Harrison, D.H. Perkins, and W.G. Scott, Phys.~Lett. B458  (1999) 79; Phys.~Lett. B530 (2002) 167.
\bibitem{LAM1} C.S. Lam, Phys.~Rev. D74 (2006) 113004; 
Phys.~Lett. B656 (2007) 193;
Phys.~Rev.~Lett. 101 (2008) 121602;
Phys.~Rev. D78 (2008) 073015.
\bibitem{LAM2} C.S. Lam, arXiv: 1104.0055.
\bibitem{LAM3} C.S. Lam, arXiv: 1105.4622.
\bibitem{SO3}  B.A. Ovrut, J.~Math.~Phys. 19 (1978) 418; G. Etesi, J.~Math.~Phys. 37 (1996) 1596; A. Adulpravitchai, A. Blum, and M. Lindner, JHEP 0909 (2009) 018; J. Berger and Y. Grossman, JHEP 1002 (2010) 071. 
\end{thebibliography}
\end{document}